# Atomic Ferris wheel beams

Vasileios E. Lembessis

*Department of Physics and Astronomy, College of Science, King Saud University, P.O. Box 2455, Riyadh 11451, Saudi Arabia*

**Abstract**

We study the generation of atom vortex beams in the case where an atomic wave-packet, moving in free space, is diffracted from a properly tailored light mask with a spiral transverse profile. We show how such a diffraction scheme could lead to the production of an atomic Ferris wheel beam.

    The advent of tunable lasers paved the way for the development of a new field in quantum optics, namely the mechanical effects of laser light on atoms and ions [1]. The coherent nature of the laser light ensures resonant interactions between an atomic particle and laser itself. This interaction involves transitions of an atomic electron among two internal energy states so we can model the atom as a two-level system. The exchange of momentum between the laser photons and the atom is the origin of changes in the atomic gross-motion. Manipulation of atomic motion led to three major application fields: optical orientation of atoms, laser selective photoionization and cooling and trapping of the atomic motion [2]. The later led to several applications in three distinct directions: optical quantum metrology, atom optics and quantum dilute gases. Today applications like Bose-Einstein condensation of atomic gases, atom optics, atom interferometry, artificial magnetism for neutral atoms, production of ultracold molecules, preparation of atomic entangled states, testing of fundamental symmetries, optical simulations of many-body physics and others are routinely produced in atomic physics labs worldwide [3]. More important is that the possibility of isolating single atoms and performing very fine experiments on them, which involve the exchange of single photons between the field and the atom, gives us the opportunity to test the fundamental postulates of quantum mechanics. So far these postulates were only tested with «gedanken experiments» , today we are able to test them experimentally thanks to the extreme degree of sensitivity in the manipulation of single atoms and photons [4].
    Two decades ago has been shown that the generation of laser beams carrying a quantized orbital angular momentum is possible [5]. Each photon of such a beam imparts a momentum as well as an angular momentum to an atom. The applications of such beams in manipulation of matter and quantum information have been rapidly growing [6], [7], [8], [9]. The mechanical effects of these beams have been extensively studied theoretically but the experimental work has not gone that far [10].
    The study of the atomic motion interacting with laser light has to take into account different parameters, like the time of interaction and the speed of atom. For example, if the interaction time is larger than the excited state relaxation time and the atomic speed is relatively large then the atomic motion can be considered as semiclassical, i.e. as the result of a radiation

pressure force which is the resultant of the so called scattering force and dipole force [2]. In the case where the atomic speed is very small the atomic de Broglie wavelength could be comparable to the laser light wavelength and the atomic particle shows a quantum behavior with a dominant wave-like behaviour. If the interaction time is very small the atomic wave-packet is diffracted after the interaction [11].

Diffraction has played a major role in the production of waves with a phase topological charge. It is the physical mechanism behind holography techniques with which we can influence the phase and/or the amplitude distribution of a light field [6]. These techniques led to the experimental realization of the Gauss-Lagurre (GL) beams which carry an orbital angular momentum $l\hbar$ per photon and have a phase singularity on the propagation axis.

Recently diffraction of electron beams has been proposed as a mechanism for producing electron vortices (EV), i.e. electron beams with an integer topological charge or quantized angular momentum along the propagation axis [12]. Vortex electrons can be created by passing a plane electron wave through spiral phase plates [13] or holographic masks [14]. This fundamentally new electron degree of freedom could find application in a number of research areas, among them the transfer of electron angular momentum to matter [15].

Soon after the the proposal and creation of EV beams it was shown that similar ideas can be used to demonstrate for the first time the possibility of creating similar beams for free atomic particles, namely the atom vortex (AV) beams [16]. The mechanism here is the diffraction of an atomic beam by a properly tailored light field (a light mask) with a fork-like intensity pattern. In that work the short interaction time between the light mask and the atom resulted to a phase imprint on the atomic wavefunction [17]. The different diffraction orders were AV beams with integer helicity traveling at different directions. The existence of such kind of beams had been theoretically proven [18], while generation of atomic vortices beams in trapped atoms has been experimentally demonstrated [19].

In this paper we have been inspired by the other pioneer experimental method for the production of vortex light and electron beams: the diffraction of a laser beam through a spiral-like phase plate [20]. We show that if we create a light mask with a spiral like intensity pattern the diffraction will give rise to AV beams. The new element here is, as it has happened with the OV beams, that the generated AV beams are focussed at different points along the beam propagation axis. By properly focussing these beams we can make them interfere. When two AV beams with opposite helicity are made to interfere then we get the atom Ferris wheel beams. These are the atomic counter-parts of the optical Ferris wheel beams with the characteristic pedal-like transverse intensity pattern [21].

A light mask with a spiral transverse profile could be taken if we interfere a GL beam with a reference light field of the same frequency of a thin round lens type. This refeerence light field can be obtained by passing an ordinary Gaussian (G) laser beam through a thin lens. We assume that both beams propagate along the $z$-direction and have a common frequency $\omega_L$ and linear polarization along the $y$-direction. The electric field of the GL beam of frequency $\omega_L$ is given by:

$$\mathbf{E}_{GL}(r,z) = E_{|\ell|,p}(r)e^{i\Theta_{\ell,p}(r,\phi,z)}e^{ikz}\hat{\mathbf{y}} \quad (1)$$

where the quantities $E_{|\ell|,p}(r,z)$ and $\Theta_{\ell,p}(r,\phi,z)$ are given by:

$$E_{|\ell|,p}(r,z) = \frac{1}{2}\sqrt{\frac{p!}{(p+|\ell|)!}}\frac{E_{GL,0}}{\sqrt{1+z^2/z_R^2}}\left(\frac{r\sqrt{2}}{w(z)}\right)^{|\ell|}\exp\left(-\frac{r^2}{w^2(z)}\right)L_p^{|\ell|}\left(\frac{2r^2}{w^2(z)}\right), \quad (2)$$

$$\Theta_{\ell,p}(r,\phi,z) = \ell\phi - (2p+|l|+1)\tan^{-1}\left(\frac{z}{z_R}\right) + \frac{kz}{2(z^2+z_R^2)}r^2. \quad (3)$$

The quantity $w(z)$ is the width of the beam, given by $w(z) = w_0\sqrt{1+z^2/z_R^2}$, with $w_0$ the beam waist and $E_{GL,0}$ the amplitude. The quantity $z_R$ is the Rayleight range of the beam given by $z_R = \pi w_0^2/\lambda$. The second term in Eq. (3) is the so called Gouy phase term while the last term is a phase shift associated with the curvature of the wave fronts. Finally the quantity $L_p^{|\ell|}(2r^2/w^2(z))$ is the associated Laguerre polynomial where the index $\ell$ is associated with the optical angular momentum of the beam ($\ell\hbar$ per photon) and the index $p$ with the characteristic intensity rings of a GL beam on the transverse plane ($p+1$ rings). The electric field of the reference Gaussian beam is given by:

$$\mathbf{E}_G = E_G(r,z)e^{i\Theta_G(r,z)}e^{-ikz}\hat{\mathbf{y}}, \quad (4)$$

where the quantities $E_G(r,z)$ and $\Theta_G(r,z)$ are given by:

$$E_G(r,z) = \frac{1}{2}\frac{E_{G,0}}{\sqrt{1+z^2/z_R^2}}\exp\left(-\frac{r^2}{w^2(z)}\right), \quad (5a)$$

$$\Theta_G(r,z) = \frac{kzr^2}{2(z^2+z_R^2)} - \tan^{-1}\left(\frac{z}{z_R}\right). \quad (5b)$$

If we consider that we pass the G beam through a thin lens of width $d$, refractive index $n$ and focal length $f$ then the electric field of the Gaussian beam is given by:

$$\mathbf{E}_G = E_G(r)e^{-iknd}e^{ikr^2/2f}e^{i\Theta_G(r,z)}e^{ikz}\hat{\mathbf{y}} \quad (6)$$

By interfering (1) and (6) we get:

$$\mathbf{E}(r,z) = e^{-ikz}e^{ikzr^2/2(z^2+z_R^2)}e^{-i\tan^{-1}(z/z_R)}\left(E_{|\ell|,p}(r,z)e^{i(\ell\phi-(2p+|\ell|)\tan^{-1}(z/z_R))} + E_G(r,z)e^{i(-knd+kr^2/2f)}\right)\hat{\mathbf{y}}$$

(7)

The intensity of the total field is proportional to the modulus squared:

$$|\mathbf{E}(r,z)|^2 = E_{|\ell|,p}^2(r,z) + E_G^2(r,z) + 2E_{|\ell|,p}(r,z)E_G(r,z)\cos\left(l\varphi - (2p+|\ell|)\tan^{-1}(z/z_R) + knd - kr^2/2f\right)$$

(8)

which at z=0 becomes:

$$|\mathbf{E}(r)|^2 = E_{|\ell|,p}^2(r) + E_G^2(r) + 2E_{|\ell|,p}(r)E_G(r)\cos\left(l\varphi + knd - kr^2/2f\right)$$

(9)

The intensity of this light field is proportional to $|\mathbf{E}(r)|^2$ and as we may see it has a spiral profile in the trasverse plane. Let's consider the following numerical example. We assume that both beams have equal beam waists $w_0 = 180\,\mu m$, equal wavelength $\lambda = 589.16$ nm and equal powers $P = 2.8$ mW. The wavelength has the value which excites the transition $6^2 S_{1/2} - 6^2 P_{3/2}$ in the $^{133}Cs$ atom. The helicity of the GL beam is $l = 2$, while the lens is characterized by the following parameter values: $n = 1.5$, $d = 0.008$ m and $f = 0.008$ m. The intensity of the total light field with the characteristic spiral transverse profile is presented in Fig.1.

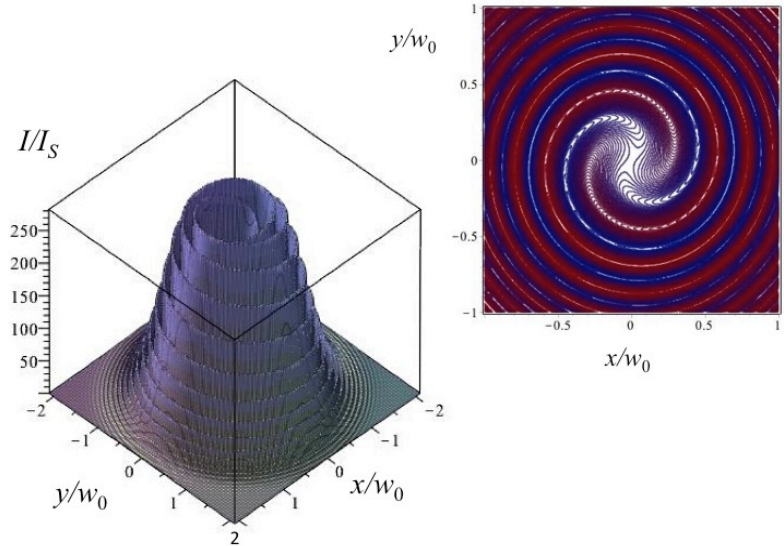

**Figure 1:** *Intensity of the total light field (at z=0) made up by the interfence of the GL beam with the G beam. The two beams have beam waists equal to $w_0 = 180\,\mu m$, a common wavelength equal to $\lambda = 589.16$ nm and the same power P = 2.8 mW. The intensity in the plot is scaled in saturation intensity units, which for the transition $6^2 S_{1/2} - 6^2 P_{3/2}$ of $^{133}Cs$ atom, is equal to $I_S = 10.9\, W/m^2$. In the inset the corresponding contour plot is shown.*

If we consider a two-level atom, with a transition frequency $\omega_0$, interacting with the above field with a Rabi frequency $\Omega(r,\phi)$ then the atom "feels" an optical dipole potential which for $|\Omega(r,\phi)|/\Delta \ll 1$ may get the form

$$U = -\frac{2\hbar|\Omega(r,\phi)|^2}{\Delta}, \quad (10)$$

where $\Delta = \omega_L - \omega_0$ is the detuning [3] and

$$|\Omega(r,\phi)|^2 = |\Omega_G(r)|^2 + |\Omega_{|\ell|,p}(r)|^2 + 2\Omega_G(r)\Omega_{|\ell|,p}(r)\cos(l\varphi + knd - kr^2/2f). \quad (11)$$

In Eq. (10) the involved quantitites are given by, $\Omega_G(r) = \Omega_{G,0}\exp(-r^2/w_0^2)$ and $\Omega_{|\ell|,p}(r) = \Omega_{GL,0}\sqrt{p!/(p+|\ell|)!}(r\sqrt{2}/w_0)^{|\ell|}\exp(-r^2/w_0^2)L_p^{|\ell|}(2r^2/w_0^2)$.

We assume the scheme shown in Fig.2. A $^{133}Cs$ atomic Gaussian wave packet moving in free space is directed towards the light mask and interacts with it for a short time interval (smaller than $\Gamma^{-1}$). The atom will be diffracted by the optical dipole potential given in Eq.(10). We consider that the atom is released from a two dimensional trap in the x-y plane. The atom was occupying the ground state of the trap and enters the interaction at time $-\tau$. The state of the atom could be described by the wavefunction $\Psi(r,-\tau) = N\exp(-4\ln 2 r^2/\sigma^2)\exp(-iK_{dB}z)$, where $K_{dB}$ is the atom's wavenumber for its motion along the z-direction. After the diffraction, the atomic wave function will get a «phase imprint» and will have the form,

$$\Psi(r,0) = \Psi_0(r,-\tau)\exp(iU\tau/\hbar) = \Psi_0(r,-\tau)\exp\left(-2i\tau|\Omega(r,\phi)|^2/\Delta\right).$$
(12)

Inserting Eq.(10) into Eq.(12), and using the Jacobi-Einger relation, $e^{iz\cos\theta} = \sum_{n=-\infty}^{\infty} i^n J_n(z)e^{in\theta}$, we get,

$$\Psi(r,\phi,0) = \Psi_0(r,-\tau)\exp(-iB\tau)\exp(-iC\tau) \times$$
$$\sum_{m=-\infty}^{\infty} i^{-m}J_m(E\tau)\exp(imknd)\exp(iml\phi)\exp(-imar^2)$$
(13)

where, $a = k/2f$, $B = 2\Omega_{|\ell|,p}^2(r)/\Delta$, $C = 2\Omega_G^2(r)/\Delta$, and $E = 4\Omega_G(r)\Omega_{|\ell|,p}(r)/\Delta$ and $J_m$ is the m-th order Bessel function . The diffraction pattern in the case

of a spiral-like potential is simpler (less rich) than in the case of a fork-like potential [16]. The main advantage, for our purpose, which is the generation of atom Ferris wheel beams, is that for the spiral apertures, the different AV beams are in focus in different planes along the propagation direction, while in the fork-like mask case they propagate in different directions.

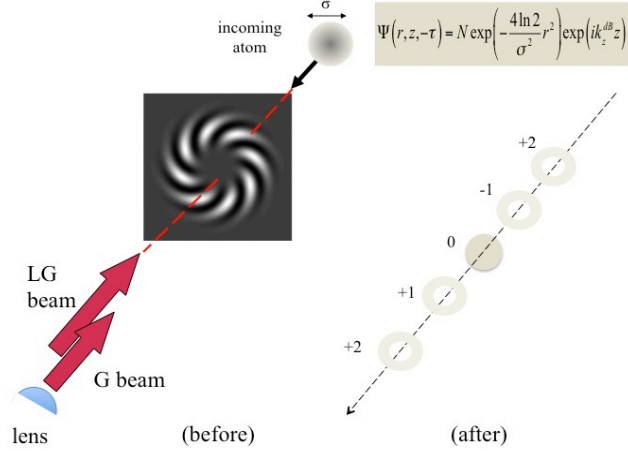

**Figure 2:** *a) Schematic representation of the diffraction of the atoms through the light mask made up of a Laguerre-Gaussian beam interfering with a Gaussian beam l = 2; (b) after the diffraction process the different atom vortices are focused at different planes along the propagation axis and are labeled m = 0, ±1, ±2, . . . with the m-th vortex carrying an orbital angular momentum equal to $m\hbar$.*

As we see from Eq.(13) the diffraction pattern is made up by a term $\Psi_0$ with no helicity and different diffraction orders of opposite helicities $\Psi_{\pm m}$ which correspond to quantized orbital angular momentum $\pm m\hbar$ along the propagation axis.

$$\Psi_0 \propto J_0(E\tau),$$

$$\Psi_m \propto -i\Psi(\phi,-\tau)\exp(-iB\tau)\exp(-iC\tau)J_1(E\tau)\exp(imknd)\exp(iml\varphi)\exp(-imar^2),$$

$$\Psi_{-m} \propto -i\Psi(\phi,-\tau)\exp(-iB\tau)\exp(-iC\tau)J_1(E\tau)\exp(-imknd)\exp(-iml\varphi)\exp(+imar^2).$$
(14)

These AV beams, described by $\Psi_{\pm m}$, are focussed at different positions due to the term $\pm mar^2$ in their phase. The spatial displacement of the different focusing is controlled by the parameter *a*. The two wavepackets with opposite angular momentum are defocused over +*mf* and –*mf* respectively.

We are going to show now how we can get atom Ferris wheel beams with our scheme. What we need is to focus AV with opposite helicities at the same points. Assume now that we send another light field to interact with the wave packets for a short time and to imprint a phase on the atom wavefunction after diffraction. Indeed the new wavefunction will be $\Psi(r,\varphi,\Delta t) \propto \Psi(r,\phi,0)\exp(-iU'\Delta t/\hbar)$, where the potential $U' = -\hbar\Omega^2/\Delta$ is the relevant optical dipole potential while Ω is the corresponding Rabi frequency associated with the new interaction. But, as we have seen, the wave function, after the diffraction from the light mask, has been splitted in different diffracting orders having different angular momenta. Assume that the light field which is used for the second phase imprint has a frequency ω and carries an optical angular momentum equal to $s\hbar$ per photon ($s$ can be any integer). When a two-level atom interacts with an optical vortex then the detuning has a Doppler shift which is characterized by mainly two contributions the axial $kV_z$ and the azimuthal $sV_\varphi/r = sL_{atom}/Mr^2$ with $L_{atom}$ the orbital angular momentum of the atom which $L_{atom} = m\hbar$ for any of the generated atom vortex beams [22]. Thus we could say that each atom vortex experiences a detuning: $\Delta = \Delta_0 - \dfrac{sm\hbar}{Mr^2}$, with $\Delta_0 = \omega - \omega_0 - kV_z$. Assume now that we consider two counter-rotating atom vortices after diffraction with angular momenta $L_{atom}^\pm = \pm|m|\hbar$ and assume that $s=1$, then the two AV will experience detunings: $\Delta_\pm = \Delta_0 \mp \dfrac{|m|\hbar}{Mr^2}$. If by proper choice of the parameters we make $\Delta_+ << \Delta_-$ then the phase imprint will be considerable only for the AV with helicity $m=1$. By arranging the phase imprint to be equal to $2mar^2$ then the two oppositely rotating AV will have a *common* phase $\exp(imar^2)$. This arrangement of the phase imprint by a light field produced by a computer generated intensity hologram [23]. Specifically it can be generated by a superposition of two vortex beams, with opposite values of $s$ and opposite radius of curvature. These two components will interfere and give us the following wave function:

$$\Psi_{\pm m} \propto -2i\Psi(\phi,-\tau)\exp(-iD\tau)\exp(-iB\tau)J_1(E\tau)\exp(+imar^2)\cos(ml\varphi + mknd) \quad (15).$$

This is clearly an atom Ferris wheel beam with a probability density $|\Psi_{\pm m}|^2 \propto 4\Psi^2(\phi,-\tau)J_m^2(E\tau)\cos^2(ml\varphi + mknd)$. Let us consider the case where the atomic Gaussian wave-packet has a standard deviation equal to $\sigma = 100$ $\mu$m. The atom is assumed to interact with the light mask for a time equal to $\tau = 0.5\Gamma^{-1}$. The relevant Rabi frequencies are $\Omega_{G,0} = \Omega_{GL,0} = 10\Gamma$ while the detuning of the first phase imprint is $\Delta = 100\Gamma$. In this case the atom Ferris wheel beam described by Eq.(17) will result in a probability density given in Fig. 3 if we assume $m = \pm 1$.

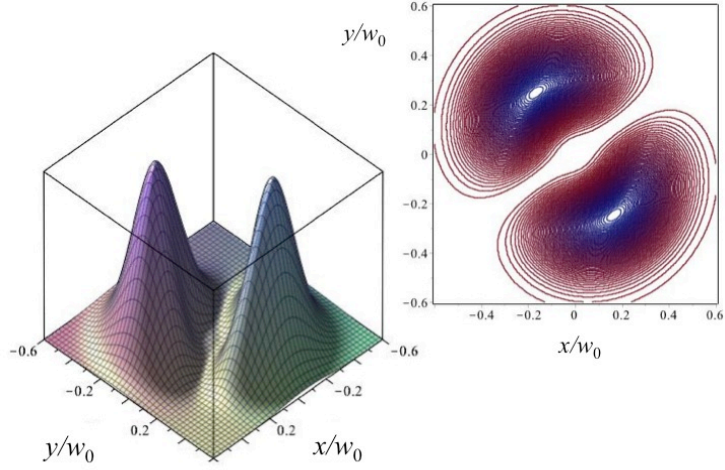

**Figure 3:** *The transverse probability distribution (arbitrary values) for the atomic Ferris wheel generated by the interference of the diffraction orders m = ±1. The time of the interaction of the atom with the light mask is $0.5\Gamma^{-1}$. In the inset the corresponding contour plot is shown.*

Following the same considerations we can create an atom Ferris wheel beam if we interfere the second order terms of Eq.(13), i.e the terms for which *m* = ±2. In this case the probability density is given in Fig.4. From both figures we see the characteristic 2ℓ pedal-like regions akin to the regions of high ligh intesnity in an optical Ferris wheel field.

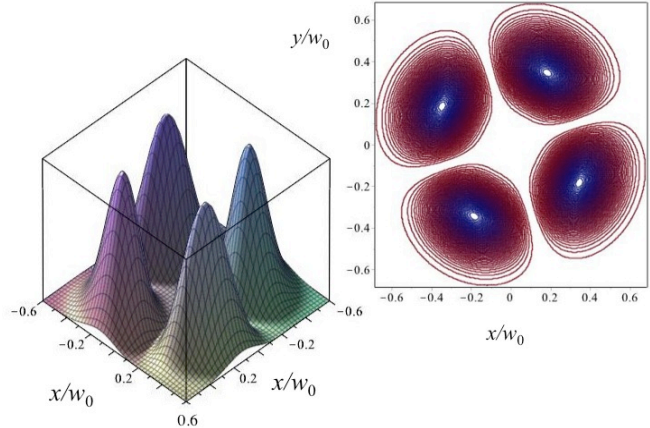

**Figure 4:** *The transverse probability distribution (arbitrary values) for the atomic Ferris wheel generated by the interference of the diffraction orders m = ±2. The time of the interaction of the atom with the light mask is $0.5\Gamma^{-1}$. $0.5\Gamma^{-1}$. In the inset the corresponding contour plot is shown.*

Finally we must point out that the diffraction process described above is a Raman-Nath diffraction [24]. There are two criteria for the validity of this type of atomic diffraction. (i) The width of the initial atomic beam must be larger than the spatial extent of the diffracting potential. This is satisfied with our choice of parameters. The transverse width of our wave-packet is $\sigma = 100$ μm while as we can see from the inset in Fig.1 the spiral character of the diffracting light field is strong within a region of less than $0.5w_0 = 90 \ \mu$m. (ii) The transverse kinetic energy of the atoms as they enter the diffraction region should be smaller than the maximum energy of the atom-light interaction. This is also satisfied for our parameters since the transverse kinetic energy is of the order of $10^{-36} J$ while the maximum energy of atom-mask interaction is of the order of $10^{-27} J$.

We have demonstrated that the diffraction of an atomic beam through a properly tailored light-field which has a spiral-like intensity pattern may produce an atomic beam which contains diffraction orders with a quantized angular momentum along the propagation direction. By a proper choise of parameters and a second phase imprint we can make AV with opposite helicities to interfere and create atom Ferris wheel type beams. In these beams the probability density forms pedal-like maxima areas like the intensity of the optical Ferris wheel beams. I thank Dr. J. Courtial for stimulating discussions. This project was funded by the National Plan for Science, Technology and Innovation (MAARIFAH), King Abdulaziz City for Science and Technology, Kingdom of Saudi Arabia, Award No. 15-MAT5110-02.